\documentstyle[12pt]{article}
\catcode`\@=11

\@addtoreset{equation}{section}

\def\@normalsize{\@setsize\normalsize{15pt}\xiipt\@xiipt
\abovedisplayskip 14pt plus3pt minus3pt%
\belowdisplayskip \abovedisplayskip
\abovedisplayshortskip  \z@ plus3pt%
\belowdisplayshortskip  7pt plus3.5pt minus0pt}
\def\small{\@setsize\small{13.6pt}\xipt\@xipt
\abovedisplayskip 13pt plus3pt minus3pt%
\belowdisplayskip \abovedisplayskip
\abovedisplayshortskip  \z@ plus3pt%
\belowdisplayshortskip  7pt plus3.5pt minus0pt
\def\@listi{\parsep 4.5pt plus 2pt minus 1pt
            \itemsep \parsep
            \topsep 9pt plus 3pt minus 3pt}}

\def\underline#1{\relax\ifmmode\@@underline#1\else
        $\@@underline{\hbox{#1}}$\relax\fi}
\@twosidetrue
\relax

\catcode`@=12

\catcode`\@=11

\def\section{\@startsection{section}{1}{\z@}{3.5ex plus 1ex minus
   .2ex}{2.3ex plus .2ex}{\large\bf}}

\def\ps@headings{\def\@oddfoot{}\def\@evenfoot{}
\def\@oddhead{\hbox{}\hfill
        \makebox[.5\textwidth]{\raggedright\ignorespaces --\thepage{}--
        \hfill }}
\def\@evenhead{\@oddhead}
\def\subsectionmark##1{\markboth{##1}{}}
}

\ps@headings

\catcode`\@=12

\relax

\def\figcap{\section*{Figure Captions\markboth
        {FIGURECAPTIONS}{FIGURECAPTIONS}}\list
        {Fig. \arabic{enumi}:\hfill}{\settowidth\labelwidth{Fig. 999:}
        \leftmargin\labelwidth
        \advance\leftmargin\labelsep\usecounter{enumi}}}
 \relax
\def\tablecap{\section*{Table Captions\markboth
        {TABLECAPTIONS}{TABLECAPTIONS}}\list
        {Table \arabic{enumi}:\hfill}{\settowidth\labelwidth{Table 999:}
        \leftmargin\labelwidth
        \advance\leftmargin\labelsep\usecounter{enumi}}}
 \relax
\def\reflist{\section*{References\markboth
        {REFLIST}{REFLIST}}\list
        {[\arabic{enumi}]\hfill}{\settowidth\labelwidth{[999]}
        \leftmargin\labelwidth
        \advance\leftmargin\labelsep\usecounter{enumi}}}
 \relax

\catcode`\@=11

\def\ps@headings{\def\@oddfoot{}\def\@evenfoot{}
\def\@oddhead{\hbox{}\hfill
        \makebox[.5\textwidth]{\raggedright\ignorespaces --\thepage{}--
        \hfill }}
\def\@evenhead{\@oddhead}
\def\subsectionmark##1{\markboth{##1}{}}
}

\ps@headings

\relax

\def\firstpage#1#2#3#4#5#6{
\begin{document}
\begin{titlepage}
\nopagebreak
\title{\begin{flushright}
        \vspace*{-1.8in}
        {\normalsize hep-th/9604062}\\[-4mm]
     {\normalsize CPTH--PC445.0496}\\[-4mm]
        {\normalsize April 1996}\\[4mm]
\end{flushright}
{#3}}
\author{\large #4 \\[1.0cm] #5}
\maketitle
\vskip -7mm     
\nopagebreak 
\begin{abstract}
{\noindent #6}
\end{abstract}
\vfill
\begin{flushleft}
\rule{16.1cm}{0.2mm}\\[-3mm]
$^{\star}${\small Talk presented at the CERN Workshop on ``STU-Dualities
and Non-Perturbative Phenomena in Superstrings and Supergravity",
27 November - 1 December 1995, and at the 2nd US-Polish Workshop
on ``Physics from Planck Scale to Electroweak Scale", Warsaw, 28-30 March
1996.}\\ 
$^{\dagger}${\small Laboratoire Propre du CNRS UPR A.0014.}
\end{flushleft}
\thispagestyle{empty}
\end{titlepage}}
\newcommand{\dal}{\raisebox{0.085cm}
{\fbox{\rule{0cm}{0.07cm}\,}}}
\newcommand{\dt}{\partial_{\langle T\rangle}}
\newcommand{\dtbar}{\partial_{\langle\bar{T}\rangle}}
\newcommand{\al}{\alpha^{\prime}}
\newcommand{\mst}{M_{\scriptscriptstyle \!S}}
\newcommand{\mpl}{M_{\scriptscriptstyle \!P}}
\newcommand{\dv}{\int{\rm d}^4x\sqrt{g}}
\newcommand{\lv}{\left\langle}
\newcommand{\rv}{\right\rangle}
\newcommand{\ph}{\varphi}
\newcommand{\abar}{\bar{a}}
\newcommand{\sbar}{\,\bar{\! S}}
\newcommand{\xbar}{\,\bar{\! X}}
\newcommand{\fbar}{\,\bar{\! F}}
\newcommand{\zbar}{\bar{z}}
\newcommand{\dbar}{\,\bar{\!\partial}}
\newcommand{\tbar}{\bar{T}}
\newcommand{\taubar}{\bar{\tau}}
\newcommand{\ubar}{\bar{U}}
\newcommand{\ybar}{\bar{Y}}
\newcommand{\phb}{\bar{\varphi}}
\newcommand{\cm}{Commun.\ Math.\ Phys.~}
\newcommand{\pr}{Phys.\ Rev.\ D~}
\newcommand{\pl}{Phys.\ Lett.\ B~}
\newcommand{\ibar}{\bar{\imath}}
\newcommand{\jbar}{\bar{\jmath}}
\newcommand{\np}{Nucl.\ Phys.\ B~}
\newcommand{\F}{{\cal F}}
\renewcommand{\L}{{\cal L}}
\newcommand{\A}{{\cal A}}
\newcommand{\e}{{\rm e}}
\newcommand{\be}{\begin{equation}}
\newcommand{\en}{\end{equation}}
\newcommand{\gsi}{\,\raisebox{-0.13cm}{$\stackrel{\textstyle
>}{\textstyle\sim}$}\,}
\newcommand{\lsi}{\,\raisebox{-0.13cm}{$\stackrel{\textstyle
<}{\textstyle\sim}$}\,}
\date{}
\firstpage{3118}{IC/95/34}
{\large\bf DUAL $N=2$ SUSY BREAKING$^{\star}$} 
{I. Antoniadis$^{\,a}$
and T.R. Taylor$^{\,a,\, b}$}
{\normalsize\sl
$^a$Centre de Physique Th\'eorique, Ecole Polytechnique,$^\dagger$
{}F-91128 Palaiseau, France\\
\normalsize\sl $^b$Department of Physics, Northeastern
University, Boston, MA 02115, U.S.A.}
{We discuss spontaneous supersymmetry breaking in $N=2$ globally
supersymmetric theories describing abelian vector multiplets. The most general
form of the action admits, in addition to the usual Fayet-Iliopoulos term, a
magnetic Fayet-Iliopoulos term for the auxiliary components of dual vector
multiplets. In a generic case, this leads to a spontaneous breakdown of one of
the two supersymmetries. In some cases however, dyon condensation restores
$N=2$ SUSY vacuum. This talk is based on the work done in collaboration with H.
Partouche \cite{apt}.}

Recently there has been revived interest in $N=2$ supersymmetry, 
in particular
in the effective actions describing non-perturbative dynamics of non-abelian
gauge  theories. In general, these theories exist only in the Coulomb phase,
with a number of abelian vector multiplets and possibly hypermultiplets, 
and their low energy effective actions can be determined exactly by using 
the underlying duality symmetries \cite{sw}.  These exact solutions
rely heavily on the restrictions following from the general structure
of $N=2$ SUSY Lagrangians.  In all known examples,
non-perturbative effects preserve $N=2$ SUSY, therefore  massless 
vector multiplet interactions are fully described by the standard
prepotential. However, a general $N=2$ SUSY theory admits also some 
Lagrangian terms that lead to spontaneous breakdown of one or both 
supersymmetries.

Only one mechanism, based on $N=1$ supersymmetric Fayet-Iliopoulos (FI)
term \cite{fay} has been known so far to break
$N=2$ supersymmetry.
It can be realized  in the presence of a $N=2$ vector multiplet
associated to an abelian gauge group factor. Decomposed under $N=1$
supersymmetry, such a  multiplet contains one vector and one chiral multiplet. 
A FI term is also equivalent to a superpotential 
which is linear in the chiral superfield. No other superpotential
seemed to be allowed for chiral components of $N=2$ vector multiplets.

Since we are interested in $N=2$ SUSY theories
viewed as low-energy realizations of some more complicated
physical systems, we do not impose the renormalizability requirement and
consider the most general form of the Lagrangian.
The basic points of our analysis can be explained on the simplest example 
of $N=2$ supersymmetric gauge theory with one abelian vector multiplet $A$
which contains besides the $N=1$ gauge multiplet $(\A_{\mu},\lambda)$
a neutral chiral superfield $(a,\chi)$.
For the sake of clarity, we begin with $N=1$ superfield description and 
rederive our results later on by using the full $N=2$ formalism.
In the absence of superpotential and FI term, the most
general Lagrangian describing this theory is determined by the analytic
prepotential
$\F(A)$, in terms of which the K\"ahler potential $K$ and the gauge
kinetic function $\tau$ are given by:
\be
K(a,\abar)={i\over 2}(a{\bar\F}_{\abar}-\abar \F_a)\qquad\qquad
\tau (a)=\F_{aa}\ ,
\label{Kf}
\en
where the $a$ and $\abar$ subscripts denote derivatives with respect to $a$ and
$\abar$, respectively. In $N=1$ superspace, the Lagrangian is written as:
\be
{\cal L}_0=-\frac{i}{4}\int d^2\theta\tau{\cal W}^2+c.c. +\int d^2\theta
d^2{\bar\theta}K
\label{L0}
\en
where $\cal W$ is the standard gauge field strength superfield.\footnote{We use 
the conventions of ref.\cite{wb}.}

The Lagrangian ${\cal L}_0$ can be supplemented by a FI term
which is linear in the auxiliary $D$ component of the gauge vector multiplet:
\be
\L_D=\sqrt{2}\xi D\ , \label{LD}
\en
with $\xi$ a real constant. It is well known that such a term preserves
also $N=2$ supersymmetry \cite{fay}.  

The Lagrangian ${\cal L}_0$ can also be supplemented by a superpotential term
\be
\L_W=\int d^2\theta\, W+c.c.
\label{LW}
\en
In order to determine what form of the superpotential
is compatible with $N=2$ supersymmetry we will impose the
constraint that the full Lagrangian,
\be
\L=\L_0+\L_D+\L_W\ ,
\label{L}
\en
be invariant under the exchange of the gaugino $\lambda$ with the fermion
$\chi$. This condition is necessary for the global $SU(2)$ symmetry under which
$(\chi,\lambda)\equiv(\lambda_1,\lambda_2)$ transforms as a doublet. It is
easy to see that it is satisfied for \cite{apt}
\be
W=ea+m\F_a\ ,\label{W}
\en
up to an irrelevant additive constant. Here $e$ and $m$ are arbitrary real
numbers. For $m=0$ the above superpotential is equivalent to a FI
term (\ref{LD}) with $\xi=e$ \cite{fay}.

After eliminating the auxiliary fields, $\L_D+\L_W$ gives rise to only two
modifications in the original Lagrangian $\L_0$. It induces the fermion mass
terms mentioned before, $\frac{1}{2}{\cal M}_{ij}\lambda_i\lambda_j$, with
\be
{\cal M}={i\over 2}\tau_a\pmatrix{e+m\taubar&i\xi\cr
i\xi&e+m\taubar}
\label{fmasses}
\en
and the scalar potential
\be
V_{N{=}1}={|e+m\tau|^2+\xi^2\over\tau_2}\ ,
\label{V1}
\en
where $\tau =\tau_1+i\tau_2$.  

In order to prove that the full Lagrangian (\ref{L}) is indeed invariant
under $N=2$ supersymmetry, we will rederive it by using the $N=2$ superspace
formalism. In this formalism, $N=2$  vector multiplets are described
by reduced chiral superfields. The reducing constraint \cite{red}
\be
(\epsilon_{ij}D^i\sigma_{\mu\nu}D^j)^2A=-96\Box A^*
\label{red}
\en
eliminates unwanted degrees of freedom, in particular by imposing 
the Bianchi identity for the gauge
field strength.  The auxiliary components of $A$,
$Y_n$, $n=1,2,3$, which form an $SU(2)$ triplet 
$\vec{Y}$,\footnote{These are related to the standard $N=1$ 
auxiliary components $F$ and $D$ by
$Y_1+iY_2=2iF, ~Y_3={\sqrt 2}D$.}
are also constraint by eq.(\ref{red}):
\be
\Box\vec{Y}=\Box\vec{Y}^*
\en
The above equation imposes real $\vec{Y}$, modulo a
constant imaginary part: 
\be
\vec{Y}=\makebox{Re}\vec{Y}+2i\vec{M}\ ,\label{YM}
\en
where $\vec{M}$ is an arbitrary real constant vector.

In terms of the reduced chiral superfield $A$, the
Lagrangian $\L_0$ can be written as
\be
\L_0={i\over 4}\int d^2\theta_1 d^2\theta_2 \F(A) + c.c.
\label{L02}
\en
As in the $N=1$ case $\L_0$ can be supplemented with a Fayet-Iliopoulos term.
Under $N=2$ supersymmetry transformations, the auxiliary components
of reduced vector multiplets transform
into total derivatives. Hence a FI term, linear in $\vec Y$, 
can be added to the action:
\be
\L_{FI}={1\over 2} {\vec E}\cdot{\vec Y} + c.c.,\label{LD2}
\en
where $\vec{E}$ is an arbitrary (complex) vector.

In order to make contact
with the $N=1$ Lagrangian (\ref{L}), we perform an $SU(2)$ transformation
which brings the parameters $\vec M$ and Re$\vec E$ into the form
\be
\vec{M}=\pmatrix{0& m &0}\qquad\qquad 
\makebox{Re}\vec{E}=\pmatrix{0& e&\xi}\ . \label{su2}
\en
It is now straightforward to show that after elimination of auxiliary fields
the Lagrangian ${\cal L}=\L_0+\L_{FI}$
coincides with (\ref{L}) up to an additive
field-independent constant. Indeed, the scalar potential is given by:
\be
V ~=~ {|\makebox{Re}{\vec E}+{\vec M}\tau |^2\over\tau_2}+
2{\vec M}\cdot\makebox{Im}{\vec E} ~=~ V_{N{=}1}+2m\makebox{Im}E_2\ .
\label{V}
\en

It can be easily shown that a non-zero parameter
$\vec M$ generates a Fayet-Iliopoulos
term for the dual magnetic $U(1)$ gauge field \cite{apt}. In fact, such a term
can be obtained from the standard electric ($\vec{M}=0$) FI term
by a duality transformation. After performing 
a symplectic $Sp(2,R)\simeq SL(2,R)$ change of basis
\be
\pmatrix{\F_a\cr a}\to\pmatrix{\alpha&\beta\cr\gamma&\delta}\pmatrix{\F_a\cr a}
\quad\qquad \tau\to{\alpha\tau+\beta\over\gamma\tau+\delta}       \label{symp}
\en
with $\alpha\delta-\beta\gamma=1$,  one obtains 
the same form of Lagrangian with new parameters $\vec M'$ and
$\vec E'$ given by
\be
\pmatrix{\vec{M}'& \makebox{Re}\vec{E}'}=
\pmatrix{\vec{M}& \makebox{Re}\vec{E}}
\pmatrix{\alpha&\beta\cr\gamma&\delta}
\qquad\quad \makebox{Im}\vec{E}'={\vec{M}\cdot
\makebox{Im}\vec{E}\over M'^2}\,\vec{M}' \ .                 \label{prime}
\en

We now turn to the minimization of the scalar potential (\ref{V}).
For $m\ne 0$, a stable minimum exists at\footnote{Without losing
generality we can choose $m,\xi\ge 0$.}
\be
\tau_1=-{e\over m}\qquad\qquad \tau_2={\xi\over m}\ .
\label{min}
\en
In this vacuum, the complex scalar
$a$ acquires the mass ${\cal M}_a=m|\tau_a|$. After diagonalizing the fermion
mass matrix (\ref{fmasses})  we find one massless fermion $(\chi-\lambda)
/\sqrt{2}$, and one massive spinor $(\chi+\lambda)/\sqrt{2}$, with the Majorana 
mass ${\cal M}_a$ equal to the scalar mass. This degeneracy is not accidental.
As we explain below, the vacuum (\ref{min}) preserves $N=1$ supersymmetry,
and the spectrum consists of one massless vector and one massive
chiral multiplets. 

In order to discuss supersymmetry breaking, it is sufficient to examine the
auxiliary field dependence of fermion transformations under $N=2$
supersymmetry:
\be
\delta\lambda_i=\frac{i}{\sqrt 2}Y_n\epsilon_{ij}(\sigma^n)^j_k\eta^k +\dots
\label{tr}
\en
where $\sigma^n$ are the Pauli matrices and the spinors $\eta^k$, $k=1,2$, are
the transformation parameters. As explained before, the effect of
the magnetic FI term amounts to introducing a constant imaginary part 
(\ref{YM}) for the auxiliary field
$\vec Y$. This constant $\makebox{Im}{\vec Y}=2{\vec M}$ enters into the
supersymmetry transformations (\ref{tr}) implying that generically both
supersymmetries are realized in a spontaneously broken mode. However, at the
minimum (\ref{min}) the real part of $\vec Y$ acquires also an expectation
value,
so that:
\be
{\vec Y} ~=~ -{2\over\tau_2}(\makebox{Re}{\vec E}+{\vec M}\tau_1)
+2i{\vec M} ~=~ 2m\pmatrix{0&i&-1}
\label{Ymin}
\en
As a result,
\be
\delta{\chi+\lambda\over \sqrt 2}=0\qquad\qquad
\delta{\chi-\lambda\over \sqrt 2}=-2im(\eta^1-\eta^2)
\label{deltas}
\en
which shows that one supersymmetry, corresponding to the diagonal
combination of the two, is preserved while the other one is spontaneously
broken. The
massless goldstino is identified as $(\chi-\lambda)/{\sqrt 2}$, in agreement 
with the spectrum found before. Hence the vector multiplet contains the
goldstino of the broken supersymmetry.

The partial breaking of extended supersymmetry might seem to contradict the
algebra of supercharges:
\be
\{{\bar Q}^i, Q_j\}=H\delta^i_j\ ,
\label{alg}
\en
where $H$ is the Hamiltonian and the spinor indices are contracted with the
metric $-{1\over 4}\sigma^0 ={1\over 4}{\bf 1}$. It follows that when one
supercharge annihilates the vacuum, $Q_1|0\rangle=0$, then the vacuum energy
vanishes and all supersymmetries remain unbroken, $Q_i|0\rangle=0$. On the other
hand if one of them is spontaneously broken then all remaining ones are
broken as
well. The loophole in this argument is that the local version of the above
algebra, which is appropriate for studying spontaneously broken symmetries, is
not the most general one \cite{pol}. The most general supercurrent algebra is:
\be
\{{\bar Q}^i,J^{\mu}_j(x)\}=T^{\mu 0}(x)\delta^i_j+\delta^{\mu 0}C^i_j\ ,
\label{ca}
\en
where $J$ is the supercurrent, $T$ is the energy-momentum tensor and $C$ is a
constant matrix. The presence of such a matrix does not affect the supersymmetry
algebra (\ref{alg}) on the fields. However some supersymmetries, namely those
associated with non-zero eigenvalues of $C$, are realized in a spontaneously
broken mode. In fact, as shown in ref.\ \cite{f2}, in the model under
consideration
\be
C^i_j=2{\vec\sigma}^i_j\cdot (\makebox{Re}{\vec E}\times{\vec M})\ .
\en

For $\xi=0$ the current algebra is not modified [$\makebox{Re}{\vec
E}\parallel{\vec M}$, see eq.(\ref{su2})] therefore partial supersymmetry
breaking does not occur. In this case, the minimum (\ref{min}) occurs at a point
where the metric $\tau_2$ vanishes. This can happen either at ``infinity''
of the
$a$-space  or at finite singular points where massless particles appear.
The quantum numbers of such states, including electric and magnetic charges,
as well as quantization conditions, depend on details of the underlying theory.
Its dynamics determines also the non-perturbative symmetries
which form a (discrete) subgroup of $Sp(2,R)$. These states cannot be vector
multiplets since unbroken non-abelian gauge group is incompatible with FI terms.
Hence we assume that they
are BPS-like dyons which form $N=2$ hypermultiplets and that the minimization
condition (\ref{min}) defines a point $a=a_0$ where one of these hypermultiplets
becomes massless. This can happen only if the parameters $(m,e)$ are
proportional to its magnetic and electric charges $(m_0,e_0)$, $(m,e)=c
(m_0,e_0)$. In order to analyze the behavior of the theory near $a_0$, one has
to include the massless hypermultiplet in the effective field theory
as a new degree of freedom. This can be done by performing the
duality transformation $A\to{\tilde A}=e_0A+m_0\F_A$, which makes possible local
description of the dyon-gauge boson interactions. In $N=1$ superspace the
superpotential (\ref{W}) becomes:
\be
W=c{\tilde a} + {\sqrt 2}{\tilde a}{\phi^+}\phi^- \ ,
\label{W0}
\en
where $\phi^{\pm}$ are the two chiral superfield components of the
hypermultiplet, and ${\tilde a}$ is the chiral component of ${\tilde A}$.

The superpotential (\ref{W0}) describes $N=2$ QED with a Fayet-Iliopoulos
term proportional to $c$ \cite{fay}.
The minimization conditions of the respective potential are
$W_{\phi^{\pm}}=0$ which is automatically satisfied at ${\tilde a}=0$ ($a=a_0$)
and
\be
W_{\tilde a}=c+{\sqrt 2}{\phi^+}\phi^-=0\ ,\qquad
{\tilde D}=0=|\phi^+|^2-|\phi^-|^2\ .
\label{Wa}
\en
As a result the dyonic hypermultiplet condenses in a $N=2$ supersymmetric
vacuum. For instance if $e=0$, the dyonic state is a pure monopole and the VEV
of the scalar field $a$ is driven to the point where the monopole becomes
massless and acquires a non-vanishing expectation value. Its condensation breaks
the magnetic $U(1)$ and imposes confinement of electric charges. This situation
is similar to the case considered in ref.\cite{sw} in the context of $SU(2)$
Yang-Mills with an explicit mass term for chiral components of gauge
multiplets which breaks $N=2$ supersymmetry explicitly to $N=1$. 

For $m=0$, the scalar potential (\ref{V}) has a runaway behavior, $V\to 0$ as
$\tau_2\to\infty$. This case is equivalent by a duality transformation to the
the case $m\neq 0$, $\xi=0$ discussed above. The runaway behavior can
be avoided if there are singular points corresponding to massless electrically
charged particles. At these points the metric
$\tau_2$ has a logarithmic singularity and the massless states have to be
included explicitly in the low energy Lagrangian to avoid non-localities. A
similar analysis of the effective theory shows that $a$ is driven then to the
points where the massless hypermultiplets get non-vanishing VEVs breaking the
$U(1)$ gauge symmetry while $N=2$ supersymmetry remains unbroken.

In the context of string theory, this phenomenon is similar to the effect
induced by a generic superpotential near the conifold singularity of 
type II superstrings compactified on a Calabi-Yau manifold \cite{s}.
In this case, the massless hypermultiplets are
black holes which condense at the conifold points. It has been shown
that such a superpotential can be generated by a VEV of the 
10-form which in four dimensions corresponds to a magnetic FI term, 
and that the black hole condensation at the conifold point leads 
to new $N=2$ type II superstring vacua \cite{ps}.

The model considered here cannot be coupled to supergravity in a straightforward
way. In particular, partial breaking of $N=2$ supergravity requires the
existence of a hypermultiplet, necessary to provide the longitudinal degrees of
freedom to the graviphoton which belongs to the massive spin 3/2 $N=1$
supermultiplet. An example of a construction leading to our model in an
appropriate globally supersymmetric limit has been given recently in ref.\
\cite{f2}.

Still in the context of global supersymmetry, it is a
very interesting question whether electric and magnetic Fayet-Iliopoulos terms
described here can be generated dynamically, for instance by an
underlying non-abelian gauge theory. It is clear that instantons do not
generate them since they give rise only to correlation functions
involving at least four fermions \cite{sei} whereas FI terms are associated
with fermion bilinears (\ref{fmasses}). Moreover, instantons respect the global
$SU(2)$ symmetry while they break the standard $U(1)$ $R$-symmetry down to
$Z_4$. On the other hand, a most general FI term violates both $SU(2)$ and
$U(1)$ leaving unbroken only a single $Z_2$, as seen from the fermion mass
matrix (\ref{fmasses}); only for $\xi=0$, $SU(2)$ remains unbroken. In general,
one cannot a priori exclude the existence of non-perturbative effects, possibly
related to gaugino condensation, which could generate FI terms in the effective
action. What seems to be most plausible is a dynamical generation of the
superpotential (\ref{W}) with $\xi=0$ which would not modify the supercurrent
algebra while merely breaking the $U(1)$ $R$-symmetry down to $Z_2$ -- a sort of
``half-instanton'' could do this job.

\vskip 0.5cm\noindent
{\bf Acknowledgments}\\
This work was supported in part by the National Science Foundation under grant
PHY--93--06906, in part by the EEC contracts SC1--CT92--0792 and
CHRX-CT93-0340, and in part by  CNRS--NSF  grant INT--92--16146.

\end{document}